\documentclass[sigconf]{acmart}
\usepackage{cleveref}

\usepackage{booktabs}
\usepackage{xcolor}
\usepackage{colortbl}
\usepackage{multirow}
\definecolor{sblue}{RGB}{230, 240, 255} 
\definecolor{sgray}{RGB}{245, 245, 245} 

\usepackage{multirow}
\usepackage{enumitem}
\usepackage{mleftright}
\usepackage{placeins}
\usepackage{tcolorbox}
\tcbuselibrary{listings}
\usepackage{booktabs} 
\usepackage{xcolor}   
\usepackage{colortbl} 
\usepackage{array}    

\renewcommand\footnotetextcopyrightpermission[1]{} 
\setcopyright{acmlicensed}
\copyrightyear{2018}
\acmYear{2018}
\acmDOI{XXXXXXX.XXXXXXX}
\acmConference[Conference acronym 'XX]{Make sure to enter the correct
  conference title from your rights confirmation email}{June 03--05,
  2018}{Woodstock, NY}
\acmISBN{978-1-4503-XXXX-X/18/06}

\begin{document}
\title{Generative Chinese Statute Retrieval}

\author{Yiteng Tu}
\affiliation{%
    \institution{Tsinghua University}
    \institution{Quancheng Laboratory}
    \city{Beijing}
    \country{China}
}
\email{yitengtu16@gmail.com}

\author{Zitao Su}
\affiliation{%
    \institution{Renmin University of China}
    \city{Beijing}
    \country{China}
}

\author{Weihang Su}
\affiliation{%
    \institution{Tsinghua University}
    \city{Beijing}
    \country{China}
}

\author{Xuanyi Chen}
\affiliation{%
    \institution{Tsinghua University}
    \city{Beijing}
    \country{China}
}

\author{Yueyue Wu}
\affiliation{%
    \institution{Tsinghua University}
    \city{Beijing}
    \country{China}
}

\author{Yiqun Liu}
\affiliation{%
    \institution{Tsinghua University}
    \city{Beijing}
    \country{China}
}

\author{Min Zhang}
\affiliation{%
    \institution{Tsinghua University}
    \city{Beijing}
    \country{China}
}

\author{Qingyao Ai}
\authornote{Corresponding Author}
\affiliation{%
    \institution{Quancheng Laboratory}
    \institution{Tsinghua University}
    \city{Beijing}
    \country{China}
}
\email{aiqingyao@gmail.com}

\renewcommand{\shortauthors}{Yiteng Tu et al.}

\begin{abstract}
Statute retrieval is a fundamental task in legal information retrieval, yet existing approaches struggle to bridge the gap between colloquial legal queries and formal statutory language.  
In this paper, we propose GCSR, a generative statute retrieval framework that reformulates statute retrieval as a sequence generation problem and internalizes statutory knowledge into a generative model. 
Specifically, we propose a multi-granularity structured docid that encodes legal hierarchy and semantic information, together with a multi-task training strategy. Experiments show that GCSR consistently outperforms strong sparse, dense, and legal-domain baselines. 
Our results demonstrate the effectiveness of generative retrieval for statute retrieval and highlight its potential for broader legal information access and downstream legal reasoning tasks.
\end{abstract}

\ccsdesc[500]{Information systems~Retrieval models and ranking}
\ccsdesc[500]{Applied computing~Law}

\keywords{Statute Retrieval, Generative Retrieval, Legal}
\settopmatter{authorsperrow=4}

\maketitle

\section{Introduction}

In modern rule-of-law societies, the rapid growth of legal information has positioned Information Retrieval (IR) as a cornerstone of legal information systems~\cite{li2023sailer,su2023caseformer,ma2023caseencoder,nguyen2024attentive,sansone2022legal}.
From legal advisory services and due diligence to judicial decision support, practitioners and citizens routinely rely on search engines to access relevant legal materials~\cite{shao2023intent,locke2022case,wiggers2022exploration}.
While statutes are foundational across legal systems, many civil-law jurisdictions such as China place particular emphasis on codified provisions as the starting point for legal authority~\cite{chen2023chinese,merryman2023civil}.
Therefore, statute retrieval, the process of identifying relevant statutory articles for specific legal queries, serves as a core component of a wide range of downstream applications, including legal question answering, legal document drafting, and public legal consultation services~\cite{su2024stard,louis2023finding}.
It is also essential to retrieval-augmented generation (RAG)~\cite{su2024dragin,su2025parametric} systems for legal applications, where retrieved statutory provisions provide authoritative grounding for model-generated outputs.
For legal professionals such as judges and lawyers, efficient retrieval systems significantly reduce the cognitive load of navigating vast legislative corpora. 
Moreover, for the general public, these systems provide a vital bridge to authoritative legal support, facilitating social justice and the accessibility of law.

However, accurately retrieving statutes remains challenging. 
On the one hand, everyday legal queries are typically formulated in colloquial, narrative language, whereas statutes are drafted using highly condensed and technical terminology~\cite{vsavelka2022legal,chen2025retrieving,yuan2025qbr}.
Bridging this mismatch requires not only lexical matching but also the ability to interpret real-world narratives and map them to legally operative facts and applicable provisions. 
For example, a query stating that “my employer did not pay me for overtime work” must be connected to legally operative issues, such as the applicable working-hours regime and the employee’s entitlement to overtime compensation, even when the query contains no corresponding statutory terminology.
On the other hand, statute retrieval necessitates complex legal reasoning that transcends simple keyword matching~\cite{louis2023finding,van2017concept}, requiring an understanding of dependencies among provisions and the hierarchy of legal principles.

Conventional legal search systems typically adopt a retrieve-then-rerank pipeline, where a first-stage retriever produces a limited candidate set and a reranker refines it~\cite{li2023thuir,xiao2021lawformer}.
Such pipelines are vulnerable to the colloquial–legal semantic gap: once the correct statute is missed at the first stage, later reranking cannot recover it, and fixed matching signals may underutilize statutory structure (e.g., hierarchy and cross-references).
In contrast, Generative Retrieval (GR) reformulates retrieval as an end-to-end sequence generation problem, directly producing document identifiers (docids) conditioned on the query~\cite{de2020autoregressive,tay2022transformer,wang2022neural}.
This paradigm is particularly attractive for statute retrieval because the corpus is relatively static and highly structured, enabling a GR model to internalize corpus semantics and decode docids that reflect statutory organization.

To this end, we propose a \textbf{G}enerative retrieval framework, GCSR, tailored for \textbf{C}hinese \textbf{S}tatute \textbf{R}etrieval.
Specifically, we first design a structured docid scheme that explicitly captures the hierarchical organization and internal structure of Chinese legislation.
It enhances the semantic richness and interpretability of identifiers compared to traditional docids.
Next, we design a multi-task training strategy encompassing general corpus training to capture document semantics, search-oriented training via pseudo-queries to bridge the colloquial-professional gap, and supervised fine-tuning on annotated query-statute pairs to establish robust relevance associations.
Experimental results show that our proposed generative approach significantly outperforms traditional sparse and dense retrieval baselines, demonstrating the promise of generative retrieval for bridging the semantic gap between user queries and statutory provisions.
In summary, this paper makes three key contributions:

\begin{itemize}[leftmargin=*]
\item  We propose a semantically and structurally grounded docid schema that improves interpretability and reduces ambiguity among similar articles.
\item We design a multi-task training strategy that integrates three complementary tasks: statute indexing, LLM-based pseudo-query augmentation, and user-query-oriented supervised fine-tuning on annotated query-statute pairs.
\item Experiments show that GCSR consistently improves over strong sparse and dense retrieval baselines on Chinese statute retrieval benchmarks.
\end{itemize}

\begin{table*}[t]
\centering
\small 
\caption{\textbf{Instantiations of the Multi-granularity Structured Docid.} The upper block presents the raw statutory input, where key spans supporting the categorization are highlighted in \textbf{bold}. The lower block displays the hierarchical semantic components generated by GCSR, transforming unstructured text into a structured, navigable retrieval path.}
\begin{tabular}{p{0.02\textwidth} p{0.14\textwidth} | p{0.33\textwidth} | p{0.43\textwidth}}
\toprule
\rowcolor{sgray}\multicolumn{2}{l}{\textbf{\textsc{Input Space}}} & \textbf{Example A (Civil Law)} & \textbf{Example B (Administrative Law)} \\
\midrule
\multicolumn{2}{l|}{\textit{Statute Title}} & 
Article 1202 of the \textbf{Civil Code} of the PRC & 
Article 24 of the Regulations on Land Compensation and Resettlement for Water and Hydropower Projects \\
\multicolumn{2}{l|}{\textit{Statute Content}} & 
Where a product defect causes harm to others, the \textbf{manufacturer} shall bear \textbf{tort liability}. & 
The relocation of \textbf{industrial and mining enterprises}... shall be compensated... restoring \textbf{original scale/standards}. \\
\midrule
\rowcolor{sblue} \multicolumn{4}{l}{\textbf{\textsc{Output Space: Structured Docid Components}}} \\
\midrule
\cellcolor{sgray}$B$ & Legal Branch & Civil and Commercial Law & Administration \\
\cellcolor{sgray}$C$ & Sub-category & Civil Code & Resources \\
\cellcolor{sgray}$T$ & Provision Type & \textbf{Substantive} (Liability Definition) & \textbf{Safeguards} (Compensation Rule) \\
\cellcolor{sgray}$L$ & Legislation Lvl. & National Law & Administrative Regulations \\
\cellcolor{sgray}$O$ & Involved Object & Producer & Industrial and Mining Enterprises \\
\cellcolor{sgray}$A$ & Abbrev. Title & Civil Code of the PRC & Large/Medium Water Conservancy Projects \\
\cellcolor{sgray}$S$ & Content Summary & Liability for tort due to product defects & Relocation/reconstruction compensation at original standards \\
\bottomrule
\end{tabular}
\label{tab:example}
\end{table*}

\section{Related Work}
\subsection{Legal Information Retrieval}
Information Retrieval (IR) has become an indispensable catalyst for legal intelligence, significantly enhancing access to justice and the efficiency of judicial proceedings by enabling the rapid identification of relevant legal sources~\cite{wiggers2023bibliometric,borgesano2025artificial,su2025judge,su2026enhancing}. 
However, existing research in legal IR has primarily focused on legal case retrieval~\cite{su2023caseformer,ma2023caseencoder,li2023thuir}, whose objective is to identify historical precedents with similar factual circumstances to ensure judicial consistency. 
Significant efforts have been directed toward modeling long-form judgment documents and capturing complex legal elements within criminal or civil cases~\cite{xiao2021lawformer,li2023sailer}. 
While these applications have advanced the automation of judicial decision-making, they primarily address analogy-based relevance between past and present cases. 
On the other hand, the retrieval of statutory articles has received relatively less attention than the retrieval of legal cases, despite being the primary source of authority in legal systems.

\subsection{Generative Retrieval (GR)}
The landscape of IR has recently undergone a paradigm shift with the emergence of Generative Retrieval (GR)~\cite{metzler2021rethinking}.
Unlike the traditional "index-then-retrieve" framework, which relies on an external inverted index or vector store, GR integrates the corpus directly into the model's parameters using a model-based indexer~\cite{metzler2021rethinking,tay2022transformer,zhou2022ultron}. 
By transforming the retrieval task into a sequence-to-sequence generation problem, in which the model directly decodes document identifiers (docids), generative approaches enable end-to-end optimization that captures deeper semantic relationships~\cite{wang2023novo,wang2022neural,zhuang2022bridging}. 
In the legal domain, the inherent hierarchy of statutes (e.g., titles, chapters, and articles) provides a natural structure for designing semantically meaningful docids. 
This paper builds upon this generative paradigm, proposing a structured DocID scheme and a multi-stage training workflow to bridge the gap between informal legal inquiries and the rigid structure of Chinese statutes.

\section{Methodology}
In this section, we present GCSR (Generative Chinese Statute Retrieval), an end-to-end framework that internalizes the statutory corpus within a generative model to bridge the semantic gap between colloquial queries and professional law. 
GCSR utilizes a multi-granularity structured docid to capture the hierarchical and functional essence of Chinese statutes, providing a semantically rich target space for decoding. 
At inference time, given a user query, GCSR decodes the top-$k$ structured docids (e.g., via beam search) and deterministically maps them back to corresponding statutory articles.
To ensure robust performance, we employ a multi-task training strategy that optimizes the model across three tasks: statute indexing, pseudo-query augmentation, and supervised retrieval.

\subsection{Multi-granularity Structured Docid}
In generative retrieval, the design of the document identifier (docid) is pivotal, as it serves as the bridge between the model's output space and the actual statutory content.
Specifically, we propose a Multi-granularity Structured docid tailored to the unique hierarchical and functional characteristics of the Chinese legal system.
Formally, for each statutory article $d$, we represent its docid as a concatenated sequence of specialized field-marked semantic tokens:
\begin{equation}
    docid(d) = [B \oplus C  \oplus T \oplus L \oplus O \oplus A \oplus S]
\end{equation}
where $\oplus$ denotes the string concatenation operator, and each component is defined as (an example is provided in~\ref{tab:example}):
\begin{itemize}[leftmargin=*]
    \item $B$ and $C$: Legal branch and sub-category. To capture the high-level legal context, we utilize a predefined legal taxonomy. 
    These tokens are identified by mapping the article to its corresponding legal department (e.g., Civil, Administrative) and sub-fields through keyword matching and structural metadata.
    \item $T$: Provision type. This captures the functional nature of the statute. We employ a rule-based classifier to identify typical lexical patterns in the text. For instance, keywords like "penalty/fine" categorize the article as a Punishment Clause, while "application/period" indicates a Procedural Clause.
    \item $L$: Legislation category. It identifies the legal hierarchy of the source (e.g., National Law, Administrative Regulation, or Judicial Interpretation) based on literal patterns in the title.
    \item $O$: Involved objects. We extract primary subjects or facilities mentioned in the text (e.g., "Industrial Enterprises," "Primary Schools"). If no specific entity is detected, a coarse-grained category is used as a placeholder.
    \item $A$: Abbreviated title. To maintain distinctiveness while ensuring brevity, we use a shortened version of the statute's title, stripping away redundant numbering.
    \item $S$: Summary. To resolve collisions between articles within the same subcategory, we employ DeepSeek-R1-Distill-Qwen-7B\footnote{\url{https://huggingface.co/deepseek-ai/DeepSeek-R1-Distill-Qwen-7B}}~\cite{guo2025deepseek} to generate a concise summary of the article's core content.
\end{itemize}

By integrating these multi-level semantic features, our multi-granularity structured docid not only enables the model to distinguish between highly similar articles but also provides a structured semantic "path" that guides the auto-regressive decoding process, significantly alleviating the mapping difficulty from colloquial queries to professional legal rules.

\subsection{Multi-task Training Strategy} \label{subsec:task}
To effectively internalize the massive statutory corpus and bridge the gap between colloquial expressions and professional legal rules, we employ a multi-task training strategy. 
This strategy optimizes the model through three complementary tasks, enabling it to function as both a robust indexer and an agile retriever:
\begin{itemize}[leftmargin=*]
\item  Statute Indexing Task. 
This task aims to encode the foundational legal knowledge into the model's parameters. 
Given the full text of a statutory article as input, the model is required to reconstruct its corresponding docid. 
By treating the statute content as a "query" for its own identifier, the model establishes a direct mapping between professional legal language and the semantic segments of the docid.
\item Pseudo-query Augmentation Task.
To alleviate the data scarcity and simulate real-world consultation scenarios, we leverage the Qwen2.5-14B model\footnote{\url{https://huggingface.co/Qwen/Qwen2.5-14B-Instruct}}~\cite{team2024qwen2} to generate 10 diverse pseudo-queries for each statutory article. 
These generated queries mimic the informal and descriptive style of non-professional users. 
Training on these \textit{(pseudo-query, docid)} pairs allows the model to learn the alignment between various linguistic variations and the authoritative legal concepts. 
\item Supervised Retrieval Task. 
Finally, to ensure the model's effectiveness in real-world applications, we fine-tune it using the authentic query-statute pairs from the official dataset. 
It forces the model to adapt to the authentic logic and noise present in real-life legal consultations, refining the decision boundaries for the most relevant docids.
\end{itemize}

In summary, the integration of these tasks creates a holistic training regime that balances the model’s ability to memorize the static legal corpus and its flexibility in understanding dynamic, colloquial user intents.
For all three tasks, we utilize a unified generative objective. 
Given an input sequence $x$ (which could be a statute text, a pseudo-query, or a real user query) and the target structured docid $y = \{y_1, y_2, ..., y_n\}$, the model is trained to minimize the negative log-likelihood via an auto-regressive objective:
\begin{equation}
    \mathcal{L} = - \sum_{t=1}^{n} \log P(y_t | y_{<t}, x; \theta)
\end{equation}
where $\theta$ represents the model parameters and $n$ is the length of the ground-truth docid sequence. 
By minimizing this loss across all tasks, the model learns to transition from various semantic spaces into the structured, hierarchical space of Chinese legal identifiers. 
This end-to-end optimization enables our generative retriever to directly optimize the docid generation rather than relying solely on similarity in embedding space, leading to more precise and context-aware statute retrieval.

\section{Experiments}
In this section, we evaluate GCSR through comprehensive experiments. We first introduce the dataset, evaluation metrics, and implementation details, then compare GCSR with representative baselines, and finally examine the effects of its structured docid design and multi-task training strategy through ablation studies.

\subsection{Datasets \& Metrics}
We evaluate our GCSR method on the STARD dataset~\cite{su2024stard}, the first Chinese statute retrieval benchmark derived from real-world legal consultations from the general public.
It comprises 1,543 non-professional query cases and a comprehensive corpus of 55,348 candidate statutory articles extracted from official Chinese laws and judicial interpretations.
We adopt Hits@$\{3, 5, 10, 20\}$ as our primary evaluation metric.
It reflects the model's ability to provide a concise and accurate set of candidate laws for practical legal consultation, where users typically only examine the top few results.

\subsection{Experimental Settings}
We employ mT5-base\footnote{\url{https://huggingface.co/google/mt5-base}}~\cite{xue2021mt5} as the backbone architecture for our GCSR model, as it provides a robust balance between multilingual representation capability and computational efficiency.
The model is trained for 300k steps using a learning rate of 1e-3 on 4 \textit{NVIDIA A100-SXM4-40GB} GPUs, with a per-device batch size of 48.
During the evaluation phase, to ensure the validity of the generated docids, the decoding strategy is constrained by a Trie (prefix tree) following previous works~\cite{tay2022transformer,de2020autoregressive}. 
We utilize beam search with a beam size of 20 to find the most probable candidate statutes for each query.

\begin{table}[t]
\centering
\setlength{\tabcolsep}{3pt}
\caption{
    \textbf{Performance comparison of different retrieval paradigms.} 
    We compare our proposed \textbf{GCSR} against traditional sparse, pre-trained, and legal retrieval models. 
    Evaluation metrics are Hits@K (H@K). 
    The best results are in {bold}, and "$*$" indicates significantly worse than GCSR at the $p < 0.05$ level using the two-tailed pairwise t-test.
}
\vspace{-2mm}
\label{tab:main}
\renewcommand{\arraystretch}{1} 
\resizebox{\columnwidth}{!}{%
\begin{tabular}{llcccc}
\toprule
\multirow{2}{*}{\textbf{Category}} & \multirow{2}{*}{\textbf{Model}} & \multicolumn{4}{c}{\textbf{Retrieval Performance}} \\ 
\cmidrule(lr){3-6}
 &  & \textbf{H@3} & \textbf{H@5} & \textbf{H@10} & \textbf{H@20} \\ 
\midrule

\multirow{2}{*}{\textit{Sparse Retrieval}} 
 & QL & 0.305* & 0.318* & 0.330* & 0.339* \\
 & BM25 & 0.319* & 0.324* & 0.337* & 0.343* \\ 
\midrule

\multirow{2}{*}{\textit{Pre-trained Embedder}} 
 & Roberta & 0.273* & 0.285* & 0.293* & 0.299* \\
 & BGE & 0.411* & 0.424* & 0.439* & 0.443* \\ 
\midrule

\multirow{2}{*}{\textit{Legal Retriever}} 
 & SAILER & 0.201* & 0.214* & 0.223* & 0.236* \\
 & Lawformer & 0.220* & 0.231* & 0.241* & 0.257* \\ 
\midrule

\multirow{2}{*}{\textit{Fine-tuned Retriever}} 
 & Dense & 0.468* & 0.486* & 0.495* & 0.507 \\
 & \cellcolor{sgray}\textbf{GCSR} & \cellcolor{sgray}\textbf{0.522} & \cellcolor{sgray}\textbf{0.536} & \cellcolor{sgray}\textbf{0.544} & \cellcolor{sgray}\textbf{0.547} \\ 
\bottomrule
\end{tabular}%
}
\vspace{-4mm}
\end{table}
\subsection{Main Results}
We compare our proposed GCSR with several representative retrieval methods, categorized into four groups:
\begin{itemize} [leftmargin=*]
    \item Sparse Retrievers. 
Traditional lexical matching methods, including QL\cite{ponte2017language} and BM25~\cite{robertson2009probabilistic}.
    \item Pre-trained Embedders. 
    Strong dense retrieval models trained on general-domain data, specifically Chinese-RoBERTa-WWM\footnote{\url{https://huggingface.co/hfl/chinese-roberta-wwm-ext}}~\cite{cui2021pre} (denoted as RoBERTa) and BGE\footnote{\url{https://huggingface.co/BAAI/bge-base-zh-v1.5}}~\cite{xiao2024c}.
    \item Legal Retrievers. 
    Models specifically pre-trained on legal corpora, including SAILER~\cite{li2023sailer} and Lawformer~\cite{xiao2021lawformer}.
    \item Fine-tuned Retrievers. 
    Following the official settings of  STARD~\cite{su2024stard}, we include a dense retriever fine-tuned on the STARD training set using RoBERTa as the backbone, denoted as Dense.
\end{itemize}

Table~\ref{tab:main} reports the performance of all models. 
First, our GCSR consistently outperforms all baseline methods across all evaluation scales. 
This demonstrates the effectiveness of the generative retrieval paradigm in the legal domain, where the model-based indexer can better internalize the authoritative knowledge of statutes compared to dual-encoder architectures.
Besides, traditional sparse retrievers (i.e., QL and BM25) surprisingly outperform several zero-shot neural models such as RoBERTa and SAILER. 
This phenomenon, also observed in the original STARD study, suggests that many statutory articles contain unique legal keywords that are crucial for matching, yet zero-shot dense models often struggle to prioritize these specific terms over general semantic similarity.
Finally, legal-domain pre-trained models (i.e., SAILER and Lawformer) show suboptimal performance, likely because these models were primarily optimized for legal case retrieval using long-form judicial documents. 
When faced with the "semantic chasm" between short, colloquial layperson queries and abstract statutory provisions, their pre-trained knowledge fails to generalize effectively.

Overall, the superior performance of GCSR highlights the advantage of our multi-task training and our structured docid design. 
By explicitly modeling the hierarchical structure of Chinese law and augmenting training with LLM-generated pseudo-queries, GCSR successfully bridges the gap between informal user intent and professional legal authority, establishing a new state-of-the-art for Chinese statute retrieval.

\begin{table}[t]
\centering
\caption{Results of the ablation experiment for the docid type. The best results are in bold.}
\begin{tabular}{c|cccc}
\toprule
docid type & Hits@3 & Hits@5 & Hits@10 & Hits@20 \\
\midrule
GCSR(ours) & \textbf{0.522} & \textbf{0.536} & \textbf{0.544} & 0.547 \\
Atomic & 0.515 & 0.528 & 0.539 & 0.541 \\
Title & 0.482 & 0.492 & 0.498 & 0.503 \\
PQ & 0.482 & 0.49 & 0.501 & 0.504 \\
Semantic & 0.517 & 0.531 & 0.542 & \textbf{0.548} \\
\bottomrule
\end{tabular}
\vspace{-2mm}
\label{tab:docid}
\end{table}
\begin{table}[t]
\centering
\caption{Results of ablation experiments for training tasks. The best results are in bold.}
\resizebox{0.99\columnwidth}{!}{
\begin{tabular}{c|c|cccc}
\toprule
Docid Type & Tasks & Hits@3 & Hits@5 & Hits@10 & Hits@20 \\
\midrule
\multirow{4}{*}{GCSR} & Full & \textbf{0.522} & \textbf{0.536} & \textbf{0.544} & \textbf{0.547} \\
 & w/o SIT & 0.486 & 0.510 & 0.516 & 0.519 \\
 & w/o PAT & 0.387 & 0.399 & 0.413 & 0.418 \\
 & w/o SRT & 0.373 & 0.380 & 0.391 & 0.395 \\
 \midrule
\multirow{4}{*}{Atomic} & Full & \textbf{0.515} & \textbf{0.528} & \textbf{0.539} & \textbf{0.541} \\
 & w/o SIT & 0.429 & 0.442 & 0.452 & 0.457 \\
 & w/o PAT & 0.386 & 0.398 & 0.408 & 0.411 \\
 & w/o SRT & 0.389 & 0.400 & 0.403 & 0.406 \\
 \bottomrule
\end{tabular}
}
\vspace{-4mm}
\label{tab:task}
\end{table}
\subsection{Ablation Study}
To evaluate the effectiveness of the various components in GCSR, we conduct a detailed ablation study from two perspectives: docid design and training task strategy.

\subsubsection{Impact of Docid Design} \hfill \break
We first investigate how different types of docids affect the performance of generative retrieval. 
We compare our proposed method with four alternatives:
\begin{itemize} [leftmargin=*]
    \item Atomic~\cite{tay2022transformer}: Assigning a unique, random integer id to each statutory article.
    \item Title~\cite{tay2022transformer}: Using the literal title of the statute as the docid.
    \item PQ~\cite{zhou2022ultron}: Using discrete codes generated by Product Quantization based on the document embeddings.
    \item Semantic: Concatenating the statute title with the summary.
\end{itemize}

The results are summarized in Table~\ref{tab:docid}. 
It can be observed that our GCSR achieves the best overall performance, indicating that integrating multi-level structural information provides stronger semantic constraints and guidance for the model. 
On the other hand, Title and PQ perform poorly. 
Titles often suffer from collisions or extreme similarity across different articles, leading to decoding ambiguity. 
PQ codes, while capturing vector features, lack human-readable semantic logic, making it harder for the model to "memorize" the mapping.
Semantic yields results close to GCSR, proving that summary information is vital for distinguishing similar articles. 
However, it still falls short of GCSR at a fine-grained level due to the lack of explicit hierarchical legal context.

\subsubsection{Impact of Training Tasks} \hfill \break
Next, we analyze the contribution of our three training tasks in \S\ref{subsec:task}, i.e, the Statute Indexing Task (SIT), the Pseudo-query Augmentation Task (PAT), and the Supervised Retrieval Task (SRT).
Table~\ref{tab:task} shows that full-task training consistently achieves the highest accuracy, and removing any single task leads to a noticeable performance decline.
Especially, PAT and SRT are more critical than SIT. 
Removing one of the two would result in a drastic performance drop. 
This confirms that fine-tuning on pseudo- or real-world judicial consultation data is effective for the model to learn the mapping between colloquial queries and professional statutes.

\section{Conclusions and Future Work}
This paper proposes GCSR, a generative statute retrieval framework that reformulates statute retrieval as a sequence-generation problem and internalizes statutory knowledge within the model parameters. 
By introducing a multi-granularity structured docid and a multi-task training strategy, GCSR effectively bridges the gap between colloquial legal queries and formal statutory language.
Experimental results demonstrate that GCSR consistently outperforms strong sparse, dense, and legal-domain baselines. 
It suggests generative retrieval as a promising paradigm for statute retrieval and can be extended to broader legal systems and downstream legal reasoning tasks.
While the proposed framework is jurisdiction-agnostic by design, our evaluation is currently limited to Chinese statutes.
We plan to apply it to other legal systems and languages. 
Additionally, accurate statute retrieval also serves as a foundational capability for downstream legal intelligence applications such as legal question answering, regulatory compliance checking, and decision support systems, which we leave for future exploration.

\bibliographystyle{ACM-Reference-Format}
\bibliography{sample-base}

@inproceedings{su2024stard,
  title={STARD: A Chinese Statute Retrieval Dataset Derived from Real-life Queries by Non-professionals},
  author={Su, Weihang and Hu, Yiran and Xie, Anzhe and Ai, Qingyao and Bing, Quezi and Zheng, Ning and Liu, Yun and Shen, Weixing and Liu, Yiqun},
  booktitle={Findings of the Association for Computational Linguistics: EMNLP 2024},
  pages={10658--10671},
  year={2024}
}

@article{cui2021pre,
  title={Pre-training with whole word masking for chinese bert},
  author={Cui, Yiming and Che, Wanxiang and Liu, Ting and Qin, Bing and Yang, Ziqing},
  journal={IEEE/ACM Transactions on Audio, Speech, and Language Processing},
  volume={29},
  pages={3504--3514},
  year={2021},
  publisher={IEEE}
}

@article{xiao2021lawformer,
  title={Lawformer: A pre-trained language model for chinese legal long documents},
  author={Xiao, Chaojun and Hu, Xueyu and Liu, Zhiyuan and Tu, Cunchao and Sun, Maosong},
  journal={AI Open},
  volume={2},
  pages={79--84},
  year={2021},
  publisher={Elsevier}
}

@inproceedings{li2023sailer,
  title={SAILER: structure-aware pre-trained language model for legal case retrieval},
  author={Li, Haitao and Ai, Qingyao and Chen, Jia and Dong, Qian and Wu, Yueyue and Liu, Yiqun and Chen, Chong and Tian, Qi},
  booktitle={Proceedings of the 46th International ACM SIGIR Conference on Research and Development in Information Retrieval},
  pages={1035--1044},
  year={2023}
}

@inproceedings{xiao2024c,
  title={C-pack: Packed resources for general chinese embeddings},
  author={Xiao, Shitao and Liu, Zheng and Zhang, Peitian and Muennighoff, Niklas and Lian, Defu and Nie, Jian-Yun},
  booktitle={Proceedings of the 47th international ACM SIGIR conference on research and development in information retrieval},
  pages={641--649},
  year={2024}
}

@article{robertson2009probabilistic,
  title={The probabilistic relevance framework: BM25 and beyond},
  author={Robertson, Stephen and Zaragoza, Hugo and others},
  journal={Foundations and Trends{\textregistered} in Information Retrieval},
  volume={3},
  number={4},
  pages={333--389},
  year={2009},
  publisher={Now Publishers, Inc.}
}

@inproceedings{ponte2017language,
  title={A language modeling approach to information retrieval},
  author={Ponte, Jay M and Croft, W Bruce},
  booktitle={ACM SIGIR Forum},
  volume={51},
  number={2},
  pages={202--208},
  year={2017},
  organization={ACM New York, NY, USA}
}

@inproceedings{xue2021mt5,
  title={mT5: A massively multilingual pre-trained text-to-text transformer},
  author={Xue, Linting and Constant, Noah and Roberts, Adam and Kale, Mihir and others},
  booktitle={Proceedings of the 2021 conference of the North American chapter of the association for computational linguistics: Human language technologies},
  pages={483--498},
  year={2021}
}

@article{team2024qwen2,
  title={Qwen2 technical report},
  author={Team, Qwen and others},
  journal={arXiv preprint arXiv:2407.10671},
  volume={2},
  number={3},
  year={2024}
}

@article{tay2022transformer,
  title={Transformer memory as a differentiable search index},
  author={Tay, Yi and Tran, Vinh and Dehghani, Mostafa and Ni, Jianmo and Bahri, Dara and Mehta, Harsh and Qin, Zhen and Hui, Kai and Zhao, Zhe and Gupta, Jai and others},
  journal={Advances in Neural Information Processing Systems},
  volume={35},
  pages={21831--21843},
  year={2022}
}

@article{zhou2022ultron,
  title={Ultron: An ultimate retriever on corpus with a model-based indexer},
  author={Zhou, Yujia and Yao, Jing and Dou, Zhicheng and Wu, Ledell and Zhang, Peitian and Wen, Ji-Rong},
  journal={arXiv preprint arXiv:2208.09257},
  year={2022}
}

@article{de2020autoregressive,
  title={Autoregressive entity retrieval},
  author={De Cao, Nicola and Izacard, Gautier and Riedel, Sebastian and Petroni, Fabio},
  journal={arXiv preprint arXiv:2010.00904},
  year={2020}
}

@inproceedings{metzler2021rethinking,
  title={Rethinking search: making domain experts out of dilettantes},
  author={Metzler, Donald and Tay, Yi and Bahri, Dara and Najork, Marc},
  booktitle={Acm sigir forum},
  volume={55},
  number={1},
  pages={1--27},
  year={2021},
  organization={ACM New York, NY, USA}
}

@inproceedings{su2025judge,
  title={Judge: Benchmarking judgment document generation for chinese legal system},
  author={Su, Weihang and Yue, Baoqing and Ai, Qingyao and Hu, Yiran and Li, Jiaqi and Wang, Changyue and Zhang, Kaiyuan and Wu, Yueyue and Liu, Yiqun},
  booktitle={Proceedings of the 48th International ACM SIGIR Conference on Research and Development in Information Retrieval},
  pages={3573--3583},
  year={2025}
}

@article{su2026enhancing,
  title={Enhancing Judgment Document Generation via Agentic Legal Information Collection and Rubric-Guided Optimization},
  author={Su, Weihang and Chen, Xuanyi and Wu, Yueyue and Ai, Qingyao and Liu, Yiqun},
  journal={arXiv preprint arXiv:2605.02011},
  year={2026}
}

@inproceedings{wang2023novo,
  title={NOVO: learnable and interpretable document identifiers for model-based IR},
  author={Wang, Zihan and Zhou, Yujia and Tu, Yiteng and Dou, Zhicheng},
  booktitle={Proceedings of the 32nd ACM International Conference on Information and Knowledge Management},
  pages={2656--2665},
  year={2023}
}

@article{wang2022neural,
  title={A neural corpus indexer for document retrieval},
  author={Wang, Yujing and Hou, Yingyan and Wang, Haonan and Miao, Ziming and Wu, Shibin and Chen, Qi and Xia, Yuqing and Chi, Chengmin and Zhao, Guoshuai and Liu, Zheng and others},
  journal={Advances in Neural Information Processing Systems},
  volume={35},
  pages={25600--25614},
  year={2022}
}

@article{zhuang2022bridging,
  title={Bridging the gap between indexing and retrieval for differentiable search index with query generation},
  author={Zhuang, Shengyao and Ren, Houxing and Shou, Linjun and Pei, Jian and Gong, Ming and Zuccon, Guido and Jiang, Daxin},
  journal={arXiv preprint arXiv:2206.10128},
  year={2022}
}

@article{su2023caseformer,
  title={Pre-training for legal case retrieval based on inter-case distinctions},
  author={Su, Weihang and Ai, Qingyao and Wu, Yueyue and Xie, Anzhe and Wang, Changyue and Ma, Yixiao and Li, Haitao and Wu, Zhijing and Liu, Yiqun and Zhang, Min},
  journal={ACM Transactions on Information Systems},
  volume={43},
  number={5},
  pages={1--27},
  year={2025},
  publisher={ACM New York, NY}
}

@article{sansone2022legal,
  title={Legal information retrieval systems: state-of-the-art and open issues},
  author={Sansone, Carlo and Sperl{\'\i}, Giancarlo},
  journal={Information Systems},
  volume={106},
  pages={101967},
  year={2022},
  publisher={Elsevier}
}

@article{ma2023caseencoder,
  title={CaseEncoder: A Knowledge-enhanced Pre-trained Model for Legal Case Encoding},
  author={Ma, Yixiao and Wu, Yueyue and Su, Weihang and Ai, Qingyao and Liu, Yiqun},
  journal={arXiv preprint arXiv:2305.05393},
  year={2023}
}

@article{nguyen2024attentive,
  title={Attentive deep neural networks for legal document retrieval},
  author={Nguyen, Ha-Thanh and Phi, Manh-Kien and Ngo, Xuan-Bach and Tran, Vu and Nguyen, Le-Minh and Tu, Minh-Phuong},
  journal={Artificial Intelligence and Law},
  volume={32},
  number={1},
  pages={57--86},
  year={2024},
  publisher={Springer}
}

@article{shao2023intent,
  title={An intent taxonomy of legal case retrieval},
  author={Shao, Yunqiu and Li, Haitao and Wu, Yueyue and Liu, Yiqun and Ai, Qingyao and Mao, Jiaxin and Ma, Yixiao and Ma, Shaoping},
  journal={ACM Transactions on Information Systems},
  volume={42},
  number={2},
  pages={1--27},
  year={2023},
  publisher={ACM New York, NY, USA}
}

@inproceedings{su2024dragin,
  title={Dragin: Dynamic retrieval augmented generation based on the real-time information needs of large language models},
  author={Su, Weihang and Tang, Yichen and Ai, Qingyao and Wu, Zhijing and Liu, Yiqun},
  booktitle={Proceedings of the 62nd Annual Meeting of the Association for Computational Linguistics (Volume 1: Long Papers)},
  pages={12991--13013},
  year={2024}
}

@article{locke2022case,
  title={Case law retrieval: problems, methods, challenges and evaluations in the last 20 years},
  author={Locke, Daniel and Zuccon, Guido},
  journal={arXiv preprint arXiv:2202.07209},
  year={2022}
}

@article{wiggers2022exploration,
  title={Exploration of domain relevance by legal professionals in information retrieval systems},
  author={Wiggers, Gineke and Verberne, Suzan and Zwenne, Gerrit-Jan and Van Loon, Wouter},
  journal={Legal Information Management},
  volume={22},
  number={1},
  pages={49--67},
  year={2022},
  publisher={Cambridge University Press}
}

@book{chen2023chinese,
  title={Chinese law: Towards an understanding of Chinese law, its nature and developments},
  author={Chen, Jianfu},
  volume={3},
  year={2023},
  publisher={Martinus Nijhoff Publishers}
}

@book{merryman2023civil,
  title={The civil law tradition: an introduction to the legal systems of Europe and Latin America},
  author={Merryman, John and P{\'e}rez-Perdomo, Rogelio},
  year={2023},
  publisher={Stanford University Press}
}

@article{louis2023finding,
  title={Finding the law: Enhancing statutory article retrieval via graph neural networks},
  author={Louis, Antoine and Van Dijck, Gijs and Spanakis, Gerasimos},
  journal={arXiv preprint arXiv:2301.12847},
  year={2023}
}

@inproceedings{chen2025retrieving,
  title={Retrieving the Right Law: Enhancing Legal Search with Style Translation},
  author={Chen, Szu-Ju and Jin, Jing and Wei, Sheng-Lun and Chen, Chien-Hung and Chen, Hsin-Hsi},
  booktitle={Proceedings of the 48th International ACM SIGIR Conference on Research and Development in Information Retrieval},
  pages={2951--2955},
  year={2025}
}

@article{yuan2025qbr,
  title={QBR: A Question-Bank-Based Approach to Fine-Grained Legal Knowledge Retrieval for the General Public},
  author={Yuan, Mingruo and Kao, Ben and Wu, Tien-Hsuan},
  journal={arXiv preprint arXiv:2505.04883},
  year={2025}
}

@article{vsavelka2022legal,
  title={Legal information retrieval for understanding statutory terms},
  author={{\v{S}}avelka, Jarom{\'\i}r and Ashley, Kevin D},
  journal={Artificial Intelligence and Law},
  volume={30},
  number={2},
  pages={245--289},
  year={2022},
  publisher={Springer}
}

@article{van2017concept,
  title={On the concept of relevance in legal information retrieval},
  author={Van Opijnen, Marc and Santos, Cristiana},
  journal={Artificial Intelligence and Law},
  volume={25},
  number={1},
  pages={65--87},
  year={2017},
  publisher={Springer}
}

@article{guo2025deepseek,
  title={Deepseek-r1: Incentivizing reasoning capability in llms via reinforcement learning},
  author={Guo, Daya and Yang, Dejian and Zhang, Haowei and Song, Junxiao and Wang, Peiyi and Zhu, Qihao and Xu, Runxin and Zhang, Ruoyu and Ma, Shirong and Bi, Xiao and others},
  journal={arXiv preprint arXiv:2501.12948},
  year={2025}
}

@article{wiggers2023bibliometric,
  title={Bibliometric-enhanced legal information retrieval: Combining usage and citations as flavors of impact relevance},
  author={Wiggers, Gineke and Verberne, Suzan and van Loon, Wouter and Zwenne, Gerrit-Jan},
  journal={Journal of the Association for Information Science and Technology},
  volume={74},
  number={8},
  pages={1010--1025},
  year={2023},
  publisher={Wiley Online Library}
}

@article{li2023thuir,
  title={Thuir@ coliee 2023: Incorporating structural knowledge into pre-trained language models for legal case retrieval},
  author={Li, Haitao and Su, Weihang and Wang, Changyue and Wu, Yueyue and Ai, Qingyao and Liu, Yiqun},
  journal={arXiv preprint arXiv:2305.06812},
  year={2023}
}

@article{borgesano2025artificial,
  title={Artificial intelligence and justice: a systematic literature review and future research perspectives on Justice 5.0},
  author={Borgesano, Francesco and De Maio, Annarita and Laghi, Pasquale and Musmanno, Roberto},
  journal={European Journal of Innovation Management},
  volume={28},
  number={11},
  pages={349--385},
  year={2025},
  publisher={Emerald Publishing Limited}
}

@inproceedings{su2025parametric,
  title={Parametric retrieval augmented generation},
  author={Su, Weihang and Tang, Yichen and Ai, Qingyao and Yan, Junxi and Wang, Changyue and Wang, Hongning and Ye, Ziyi and Zhou, Yujia and Liu, Yiqun},
  booktitle={Proceedings of the 48th International ACM SIGIR Conference on Research and Development in Information Retrieval},
  pages={1240--1250},
  year={2025}
}

\end{document}